\documentclass[twocolumn]{jpsj2}

\usepackage{graphicx}

\title{
Possibility of Unconventional Pairing Due to Coulomb Interaction 
in Fe-Based Pnictide Superconductors: Perturbative Analysis 
of Multi-Band Hubbard Models}

\author{Takuji \textsc{Nomura}\thanks{E-mail address: nomurat@spring8.or.jp}}

\inst{
Synchrotron Radiation Research Center, Japan Atomic Energy Agency, 
Sayo, Hyogo 679-5148
}

\recdate{\today}

\abst{
Possibility of unconventional pairing due to Coulomb interaction 
in iron-pnictide superconductors is studied by applying a 
perturbative approach to realistic 2- and 5-band Hubbard models. 
The linearized Eliashberg equation is solved by expanding 
the effective pairing interaction perturbatively up to third order 
in the on-site Coulomb integrals. 
The numerical results for the 5-band model 
suggest that the eigenvalues of the Eliashberg equation 
are sufficiently large to explain the actual high $T_{\rm c}$ 
for realistic values of Coulomb interaction 
and the most probable pairing state is spin-singlet 
$s$-wave without any nodes just on the Fermi surfaces, 
although the superconducting order parameter changes 
its sign between the small Fermi pockets. 
On the other hand the 2-band model is quite insufficient 
to explain the actual high $T_{\rm c}$. }

\kword{
iron-pnictide superconductors, superconducting mechanism, 
pairing symmetry}

\begin{document}
\maketitle

Recent discovery of high-$T_{\rm c}$ superconductivity 
in iron pnictides has generated highly intensive 
research activities in solid state physics. 
After the discovery of superconductivity in LaFeAsO$_{1-x}$F$_x$ 
system~\cite{ref:KamiharaY2008}, 
it has become evident that transition temperature is raised 
above 40K by replacing La with other rare-earth elements 
(Ce, Pr, Nd, Sm, ...)~\cite{ref:ChenGF2008,ref:RenZA2008a,
ref:RenZA2008b,ref:ChenXH,ref:RenZA2008c}
 or by applying pressure~\cite{ref:TakahashiH2008}. 
Pairing mechanism and symmetry would be the most intriguing 
issues of this newly discovered superconductivity. 
In the present work, possibility of unconventional 
(i.e., not phonon-mediated) superconducting mechanism 
for Fe-pnictide superconductors is investigated theoretically. 
There are already several reasons why unconventional 
pairing may be realized in iron pnictides: 
(i) $T_{\rm c}$ is high, compared with conventional 
(phonon-mediated) BCS superconductors, 
(ii) Fe3d states dominate the most part of the density 
of states near the Fermi level~\cite{ref:SinghDJ2008, 
ref:BoeriL2008, ref:KurokiK2008}, 
and (iii) electron-phonon coupling is expected 
to be weak by first-principles calculations~\cite{ref:BoeriL2008}. 

We introduce many-band Hubbard models 
for Fe3d-like orbitals and formulation. 
The Hamiltonian is given in the form $H=H_0+H'$. 
$H_0$ is the non-interacting part: 
$H_0= \sum_{ij} \sum_{\ell\ell'} \sum_{\sigma} 
t_{i\ell, j\ell'} c_{i\ell\sigma}^{\dag} c_{j\ell'\sigma}$, 
where $c_{i\ell\sigma}$ ($c_{i\ell\sigma}^{\dag}$) 
is the electron annihilation (creation) 
operator for Fe3d-like orbital $\ell$ with spin $\sigma$ at site $i$. 
The tight-binding parameters $t_{i\ell, j\ell'}$ are determined 
to reproduce a realistic electronic structure. 
$H'$ is the on-site Coulomb interaction part 
containing four kinds of Coulomb integrals: 
$U$ (intra-orbital repulsion), $U'$ (inter-orbital repulsion), 
$J$ (Hund's coupling), $J'$ (inter-band pair-hopping). 
The same form of $H'$ was used for the Ru4d$\varepsilon$-like electrons 
of Sr$_2$RuO$_4$ in Ref.~\ref{ref:NomuraT2002}. 
Then the following linearized Eliashberg equation 
is solved numerically: 
\begin{eqnarray}
\lambda \cdot \Delta_{a, \sigma_1\sigma_2}(k) &=& 
- \frac{T}{N}\sum_{k', a', \sigma_3\sigma_4} 
V_{a\sigma_1\sigma_2, a'\sigma_3\sigma_4}(k, k') \nonumber\\
&&\times |G_{a'}^{(0)}(k')|^2 \Delta_{a',\sigma_4\sigma_3}(k')
\end{eqnarray}
where $a$ and $a'$ are band indices, $G_a^{(0)}(k)$ 
is the Green's function for band $a$ (without self-energy corrections), 
$\sigma_i$'s are spin indices, 
$V_{a\sigma_1\sigma_2, a'\sigma_3\sigma_4}(k, k')$ 
is the effective pairing interaction, 
$\Delta_{a, \sigma\sigma'}(k)$ is the anomalous self-energy 
on band $a$, and $\lambda$ is eigenvalue. 
The effective pairing interaction is evaluated 
by third order perturbation expansion in $H'$. 
The third order perturbation theory has been applied to many other 
unconventional superconductors, and suggests correctly 
pairing symmetries, e.g., singlet $d_{x^2-y^2}$-wave for cuprates 
and organic superconductors, triplet $p$-wave for Sr$_2$RuO$_4$, 
... etc.~\cite{ref:YanaseY2003}
Transition point is determined by $\lambda=1$. 
We take 32$\times$32 ${\mib k}$ points and 512 Matsubara 
frequencies for numerical calculations. 

\begin{figure}
\begin{center}
\includegraphics[width=0.8\linewidth]{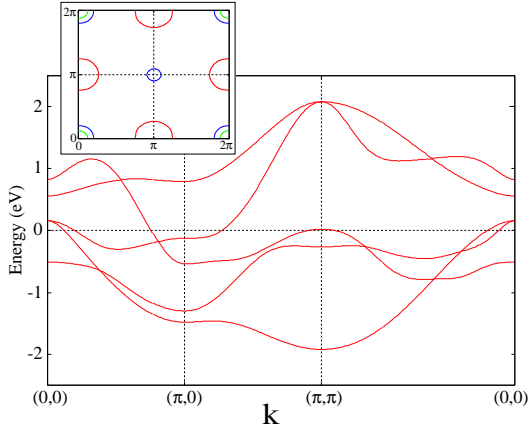}
\end{center}
\caption{
Electronic band structure for the 5-band model.
Inset shows the Fermi surface. }
\label{Fig1}
\end{figure}

\begin{figure}
\begin{center}
\includegraphics[width=0.8\linewidth]{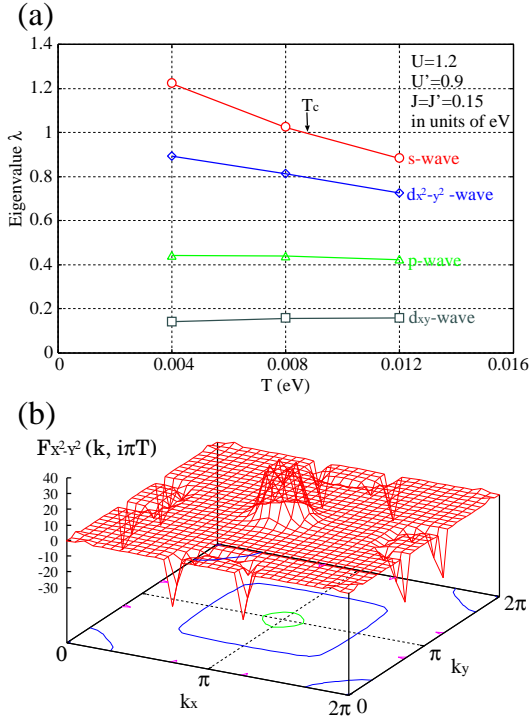}
\end{center}
\caption{
(a) Eigenvalue $\lambda$ for various pairing symmetries 
as a function of temperature in the 5-band model.
(b) Anomalous Green's function of the local 
Fe3d$_{X^2-Y^2}$-like orbitals, $F_{X^2-Y^2}({\mib k}, {\rm i} \pi T)$, 
at $T=0.008$.}
\label{Fig2}
\end{figure}

Firstly, we adopt two-dimensional 5-band tight-binding 
model proposed by Kuroki et al~\cite{ref:KurokiK2008}. 
The electronic structure and the Fermi surface are given 
in Fig.~\ref{Fig1} (in the unfolded representation, 
where each unit cell contains only one Fe atom). 
The Fermi surface consists of hole pockets 
around the $(0, 0)$ and $(\pi, \pi)$ points and electron pockets 
around the $(\pi, 0)$ and $(0, \pi)$ points. 
In the original folded representation, 
where each unit cell contains two Fe atoms, the $(0, 0)$ and $(\pi, \pi)$ points 
are folded onto the $\Gamma$ point, while the $(\pi, 0)$ and $(0, \pi)$ points 
onto the M point. 
The numerical results of eigenvalues are shown in Fig.~\ref{Fig2}(a). 
We see the most probable pairing symmetry is singlet $s$-wave 
and obtain sufficiently large eigenvalues to explain actual 
high $T_{\rm c}$'s for realistic values of Coulomb integrals 
($U=1.2{\rm eV}$, $U'=0.9{\rm eV}$, $J=J'=0.15{\rm eV}$). 
$T_{\rm c}$ is evaluated to be about 100K. 
This is still higher than real values 20K-50K. 
If we include the self-energy corrections, then $T_{\rm c}$ 
will be decreased somewhat due to the effect of quasi-particle damping. 
The d$_{X^2-Y^2}$-electron component of anomalous Green's 
function $F_{X^2-Y^2}({\mib k}, {\rm i} \omega_n)$  
is shown in Fig.~\ref{Fig2}(b) ($X$ and $Y$ axes are 
those in the original folded representation, 
while ${\mib k}=(k_x, k_y)$ is in the unfolded representation).  
The superconducting order parameter does not possess any 
nodes just on the Fermi surfaces, although it changes 
its sign between the electron and hole pockets. 
In this sense the pairing symmetry is {\it extended} $s$-wave. 

\begin{figure}
\begin{center}
\includegraphics[width=0.8\linewidth]{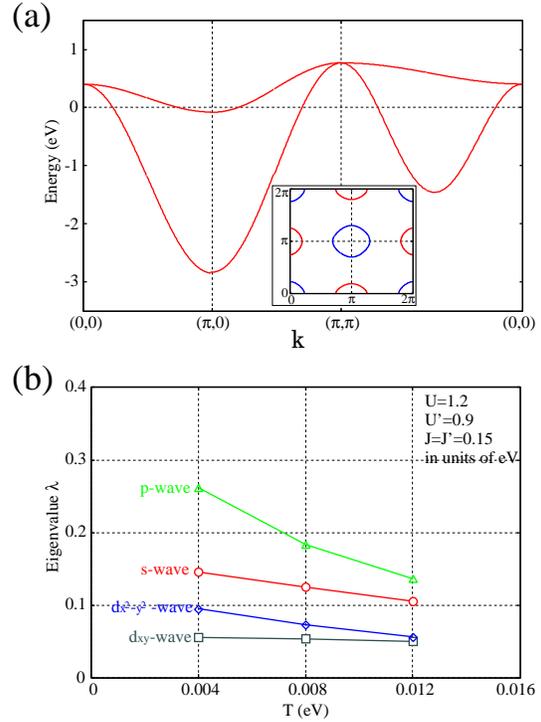}
\end{center}
\caption{
(a) Electronic band structure for the 2-band model.
Inset shows the Fermi surface. 
(b) Eigenvalue $\lambda$ for various pairing 
symmetries as a function of temperature in the 2-band model.}
\label{Fig3}
\end{figure}

We proceed to another model, i.e., 2-band model only 
for the d$_{xz}$- and d$_{yz}$-like orbitals, 
proposed by Raghu et al.~\cite{ref:RaghuS2008} (See Fig.~\ref{Fig3}(a)). 
The maximum eigenvalue is given by the triplet $p$-wave pairing state, 
but is too small to explain the actual $T_c$, as we see in Fig.~\ref{Fig3}(b). 
Thus we may conclude that the 2-band model is quite insufficient 
and the real multi-band situation is essential to describe 
the iron-pnictide high-$T_c$ superconductivity. 

In conclusion, our perturbation theory suggests 
that the iron-pnictide superconductivity may 
be unconventional one induced 
by electron correlation effect, 
as other unconventional superconductivity. 
The most probable pairing symmetry is $s$-wave 
without any nodes on the Fermi surface. 
One of the important differences from other unconventional 
superconductors is that the order parameter will change 
its sign not on the Fermi surface but between the Fermi pockets. 
Possibility of triplet pairing is excluded. 

The author would like to thank H. Ikeda, 
K. Kuroki, K. Nakamura, S. Onari and K. Yamada 
for valuable communications. 
Numerical work was performed at the Yukawa Institute 
Computer Facility of Kyoto University. 

\bibliography{99}

\end{document}